\documentclass{article}
\usepackage{PRIMEarxiv}
\usepackage[utf8]{inputenc}
\usepackage[T1]{fontenc}
\usepackage{microtype}
\usepackage{cite}
\usepackage{amsmath,amssymb,amsfonts}
\usepackage{graphicx}
\usepackage{booktabs,tabularx,array}
\usepackage{textcomp}
\usepackage{placeins}
\usepackage{float}
\usepackage[font=small,labelfont=bf,justification=raggedright,singlelinecheck=false]{caption}
\usepackage[hidelinks]{hyperref}
\newcolumntype{Y}{>{\raggedright\arraybackslash}X}
\newcolumntype{P}[1]{>{\raggedright\arraybackslash}p{#1}}
\title{Can Model Answer Probabilities Represent Survey Populations? Conditional Prediction and Joint Validity in Jev Survey Simulation}
\author{Zengqing Wu\textsuperscript{1} \quad
Chuan Xiao\textsuperscript{1,2}\thanks{Corresponding author. \href{mailto:chuanx@nagoya-u.jp}{chuanx@nagoya-u.jp}. Our source code is available at \url{https://github.com/wuzengqing001225/jev-survey-distributions}.}\\[4pt]
{\normalfont\small\textsuperscript{1}University of Osaka, Japan}\\
{\normalfont\small\textsuperscript{2}Nagoya University, Japan}}
\date{}

\makeatletter
\AtBeginDocument{\immediate\write\@auxout{\string\citation{IEEEReferenceControl}}}
\makeatother

\begin{document}
\raggedbottom
\maketitle

\begin{abstract}
Survey simulations require realistic combinations of answers, but predictive accuracy alone does not establish that a model can generate them. We evaluate the option probabilities returned by Jev, a TypeSafe decision model, using linked responses to 20 binary items from five archived survey waves. We compare population-only prediction, prediction given recorded answers, and sequential completion that uses generated answers to predict later responses. Joint comparisons hold the starting population information fixed, and two three-question cases are completed in both question orders. We evaluate predictive scores, answer frequencies, and associations against the survey records. Recorded answers can improve predictive scores while associations remain distorted. Large conditional errors also occur. For example, among men who described themselves as having traditional values, the model predicted that 16\% would see men and women as similar in workplace abilities, compared with 56\% in the survey. Sequential completion has mixed effects on joint validity, and reversing question order produces different joint distributions in both tested cases. Order checks and classical marginal bounds clarify what can be assessed before linked validation answers are available. The findings support validating the joint events a simulation uses and separating empirical constraints from model assumptions. The evidence concerns selected binary tasks and one model version.
\end{abstract}


\section{Introduction}
With the growing use of large language models, survey simulation is expanding beyond academic research into commercial market research \cite{park2024llm,simile2026}. For example, Simile offers simulated populations for product and pricing research, aiming to reduce the cost of repeated human surveys. Many applications need realistic combinations of answers. A simulation of financial vulnerability, for example, may need to estimate how many people have both credit-card debt and a car loan. Knowing each debt's prevalence does not determine their overlap. Errors in that relationship can change simulated outcomes even when answer proportions look plausible.

Numerical probability interfaces offer a direct way to construct such populations: a simulator can use the returned probabilities to assign answers and combine conditional predictions into complete records. Jev is a commercial decision model developed by TypeSafe. Its multiple-choice interface, identified by the question type \texttt{choice}, returns a selected option, a probability for each option, and a separate confidence field \cite{typesafe}. We study the option probabilities under an explicit frequency target: the recorded answer of a respondent drawn at random from a specified survey cohort. A probability for one person's unknown answer and the frequency of that answer across people are different quantities. Their correspondence requires empirical validation.

Existing survey-simulation evaluations examine opinion alignment, population distributions, associations, and individual prediction \cite{santurkar2023whose,xie2026evaluating,duan2026general}. The question here is how the use of a fixed probability interface changes between predicting a recorded answer and constructing a complete respondent. To predict an existing respondent's car-loan answer, the model may use that person's recorded credit-card-debt answer. To generate a virtual respondent, it may instead generate the debt answer and use it to generate the car-loan answer. These operations use related conditional predictions but have different validation targets. Conditional predictions that describe one common joint distribution should also give the same answer combinations when the completion order is reversed.

We use linked answers from archived American Trends Panel surveys to compare population-only prediction, recorded-answer conditioning, and completion using generated answers. Joint comparisons share the same human reference information to isolate differences between completion rules. The study addresses three research questions:
\begin{enumerate}
\item \textbf{RQ1: Do population-targeted probabilities reproduce human answer shares?}
\item \textbf{RQ2: How does an observed answer affect prediction and human association fidelity?}
\item \textbf{RQ3: How do sequential completion and completion order affect the joint distribution?}
\end{enumerate}

The results answer these questions in turn. Population-only probabilities reproduce some shares closely but miss others by up to 23.5 percentage points. Recorded answers improve predictive scores in several comparisons, yet the resulting associations can remain too strong or too weak. Using generated answers to complete later responses has mixed effects on joint validity. Reversing the order changes the generated population in both tested three-question cases, revealing internally incompatible conditional predictions.

The study contributes a controlled evaluation of the transition from single-answer prediction to population generation. It demonstrates that predictive gains, fidelity to human associations, and internal compatibility require separate evidence. It also connects these empirical distinctions to established probability bounds, clarifying which features are constrained by the available human information. For researchers constructing synthetic respondents, the findings motivate four checks: validate probabilities as population frequencies, evaluate the joint events the application uses, compare completion orders, and distinguish empirical constraints from model assumptions. The evidence concerns selected binary cases and one model version.

\section{Related Work}
Survey realism already has several evaluation targets. OpinionQA evaluates how opinions from large language models (LLMs) align with those of human demographic groups \cite{santurkar2023whose}. SocialScienceDataBench (SSDataBench) examines population statistics including distributions, associations, multivariate prediction, and sequences \cite{xie2026evaluating}. The study introducing the JEVal decision-model benchmark evaluates numerical probabilities, individual survey prediction, and population-level simulation \cite{duan2026general}. Park et al. ground individual agents in interviews and survey answers and evaluate their predictions on held-out responses \cite{park2024llm}. These works establish frequency and dependence structure as existing evaluation concerns. We focus on how the use of a fixed probability interface changes between predicting recorded answers and constructing a complete response population.

FullRespondent-LLM (FR-LLM) trains marginal and respondent models, uses generated history and sampled item orders at inference, then reweights complete response vectors toward predicted item marginals \cite{huang2026simulating}. Semantically Coherent Batching and Ordering (SCBO) shares references and orders batched answer generation without model fine-tuning \cite{li2026scbo}. These methods construct related answers using learned or prompt-based dependencies. Our controlled comparisons assess the conditional probabilities supplied by the existing probability interface. The two completion rules use the same starting answer shares and first-answer probabilities. This isolates how generated history changes the resulting joint distribution.

Williams et al. evaluate correlation structures built mainly from demographic subgroup means and discuss respondent trajectories as another evaluation object \cite{williams2026beyond}. Krsteski et al. study human-label allocation for synthesis and estimate correction, with additional individual-error and multi-turn analyses \cite{krsteski2026valid}. E2Sim, a framework for effective and efficient LLM social survey simulation, learns a model/history budget policy from accuracy and cost supervision \cite{li2026adaptive}. These approaches address representation and the use of human information. We validate the resulting answer combinations against linked individual records. The generator receives the starting answer shares, and no model is fitted to the target answers.

Sys1Cal-v1 evaluates numerical probabilities against synthetic truth known by construction \cite{porcedda2026jev}. A public Jev survey evaluation reports elicitation and demographic effects \cite{jevpublic}. Our target is the distribution of recorded human answers, including how its conditional probabilities combine into complete response patterns.

Conditional compatibility provides a statistical basis for this question. In multiple imputation, separately specified conditional models can lack a common joint distribution, and update order can affect the resulting distribution \cite{van2007multiple}. We apply this established concern to the finite joint tables induced by two completion orders. Their total-variation distance gives a check of internal compatibility and a bound on agreement with any common target population.

\section{Materials and Probability Targets}
\subsection{Selected Cases and Their Units}
We use public American Trends Panel (ATP) responses distributed through OpinionQA. The prefix W identifies a survey wave, such as W92 for Wave 92. The 13 comparisons consist of four W92 pairs, two W54 pairs, two W41 pairs, two W29 observed-answer predictions, and three triples. They involve 20 unique items in five waves. Finance1 is an exploratory comparison that reuses the W54 pair items. Panel members may recur across waves. A historical-context check uses the W29 workplace item.

Cases were selected for binary wording and the availability of answers to the required questions from the same respondents. A denotes the known starting answer, and B and C denote the later answers. Race and Finance2 use both B-to-C and C-to-B completion orders, keeping A first. Appendix~\ref{app:selection} documents selection and data reuse.

\begin{table}[tbp]
\caption{Selected case material. Counts include evaluation respondents whose required answers fall within the two listed options. Full wording and options appear in the appendix.}
\label{tab:cases}
\centering\small
\begin{tabularx}{\textwidth}{P{1.7cm}P{3.0cm}Y P{2.0cm}}
\toprule
Block & Given answer or attribute & Target answer(s) & Evaluation n \\
\midrule
W92 pairs & Economic system: unfair or fair & Corporate profit, government size, federal health coverage, labor unions & 4891--10052 \\
W54 pairs & Savings account & Retirement account, medical-bill debt & 3365, 3377 \\
W41 pairs & Race relations in 2050 & Public education, political division in 2050 & 1088, 1084 \\
W29 predictions & Recorded sex and traditional values & Similarity of men/women in parenting and workplace abilities & 1242 \\
Finance1 (W54) & Savings account & Retirement account (B), medical-bill debt (C) & 3360 \\
Finance2 (W54) & Investment ownership & Credit-card debt (B), car loan (C) & 3360 \\
Race (W43) & Perceived discrimination problem & Race relations (B), focus for improvement (C) & 1612 \\
\bottomrule
\end{tabularx}
\end{table}

The field periods are September 14--28, 2017 (W29), December 11--23, 2018 (W41), September 16--29, 2019 (W54), July 8--18, 2021 (W92), and January 22--February 5, 2019 (W43) \cite{pew29,pew41,pew54,pew92,pew43}. W41 records opinions about 2050 rather than later realized outcomes. Full item wording and option coding appear in Appendix~\ref{app:dictionary}.

Finance1 uses savings ownership as A, retirement ownership as B, and medical debt as C. Finance2 uses investment ownership outside retirement accounts as A, credit-card debt as B, and a car loan as C. Race uses the perceived discrimination problem as A, overall race relations as B, and the focus for improving relations as C.

\subsection{What the Requests Ask}
We query \texttt{jev-1.14.0} for the probability of each answer under each combination of known information. The same probability vector applies to all respondents with that information. An observed answer is a recorded human response. A generated answer is a possible response assigned a model probability. We calculate the joint distribution by enumerating every answer combination and multiplying its branch probabilities. This gives the distribution implied by the completion rule without randomly drawing virtual respondents. Survey sample sizes count the human records used for evaluation, not model requests.

 In W54, W41, W29, and W43, a seeded random assignment places half the wave's rows in a reference group and the rest in an evaluation group before eligibility filtering. In the W54, W41, and W43 joint comparisons, the reference group supplies the weighted shares of the starting answer A, $r(a)$, and the evaluation group supplies the linked answers used for scoring. Reference B/C values determine eligibility but their category frequencies are not used in prediction. The W29 comparisons use only the evaluation group. W92 uses all eligible respondents for the matched-mean diagnostic defined below, without a reference/evaluation split. The model receives hypothetical combinations of known answers, item wording, answer options, and a cohort description.

\noindent\begin{minipage}{\columnwidth}
The following example instruction defines the prediction target:
\par\smallskip
\begingroup
\setlength{\fboxsep}{6pt}
\setlength{\fboxrule}{0.3pt}
\noindent\fbox{\begin{minipage}{\dimexpr\columnwidth-2\fboxsep-2\fboxrule\relax}
\small
\textbf{Example prediction instruction}\par\smallskip
A respondent is drawn at random from population, conditional on all \texttt{known\_answers} in \texttt{state}. Each known answer was given by this same respondent. Which response did this respondent give to \texttt{survey\_question}? Predict the respondent's recorded choice, including disagreement, rather than choosing the most correct or socially desirable answer.
\end{minipage}}
\endgroup
\par\smallskip
\end{minipage}
\par\smallskip
For this request, the state names ``Pew Research Center American Trends Panel, archived Wave54.'' The known-answer record contains the savings question, its ``Yes, have this'' response, and \texttt{FINANCEa\_W54}. The target instruction contains \texttt{FINANCEb\_W54} and the retirement-account wording. The full requests are included in the reproduction package.

The target is the weighted distribution of historical answers in the eligible survey cohort. The historical-context check asks whether explicitly identifying that period changes W29 workplace prediction. It adds both the field-period dates and an instruction to predict answers from that period. Source-identifier controls replace survey names and item codes with neutral aliases while preserving question wording, options, respondent information, eligibility, and dates within each comparison. The identifier controls specify September 2017 for W29 and September 2019 for W54, with the period fixed across named and masked conditions.

\section{Evaluation Design and Statistical Analysis}
\subsection{Observed Answers and Joint Completion}
Let $u(b)$ be a population-only prediction and $q(b|a)$ a prediction given observed A. With the same reference marginal,
\begin{equation}
J_0(a,b)=r(a)u(b),\qquad J_1(a,b)=r(a)q(b|a).
\end{equation}
The population-only control measures the value of using A under the same supplied A shares. A matched-mean control uses $m(b)=\sum_a r(a)q(b|a)$. For the W54/W41 joint comparisons, $K(a,b)=r(a)m(b)$ has the same A and B margins as $J_1$, with no association between them. Comparing K with $J_1$ measures the prediction gain from their different associations at fixed margins. W92 assigns the same mean forecast to everyone and compares it with forecasts given each person's A answer, using A shares from the evaluation records as a within-cohort diagnostic. W29 asks whether a known answer adds information beyond a demographic attribute: it compares $q(B|X)$ with $q(B|X,A)$, where X is recorded sex and A is the traditional-values answer.

For triples, separate completion and sequential completion are
\begin{align}
J_{\rm ind}(a,b,c)&=r(a)q_B(b|a)q_C(c|a),\\
J_{\rm seq}(a,b,c)&=r(a)q_B(b|a)q_C(c|a,b).
\end{align}
Sequential completion uses the generated B answer as input when predicting C. The separate and sequential rules share the A marginal and the first generated-answer probabilities. We enumerate all eight patterns and compare their distribution with the weighted human records. Reverse completion uses $J_{\rm rev}(a,b,c)=r(a)q_C(c|a)q_B(b|a,c)$. Order comparisons evaluate both directions together against a common separate-completion baseline. Each source-identifier condition uses its own baseline. Absolute joint errors refer to the main completion comparisons. Order and source-identifier effects are estimated within their respective comparisons.

\subsection{Evaluation Measures}
The main comparison measure is the unhalved categorical Brier score, which measures squared probability errors against observed answer categories,
\begin{equation}
L_B(p;H)=1+\sum_k p_k^2-2\sum_k H_kp_k,
\end{equation}
where H is the weighted empirical pattern distribution. Gain is control loss minus treatment loss, so a positive gain favors the treatment. The score ranges from 0 to 2 and is interpreted within each task because outcome frequencies and category counts affect reference losses \cite{gneiting2007strictly}.

We also report joint total variation (TV), $\mathrm{TV}(p,H)=\frac12\sum_k|p_k-H_k|$, target-share errors, conjunction errors, and association strength. TV is the minimum probability mass that must be reassigned to make two distributions match, reported as a percentage. Binary share differences are reported in percentage points (pp). Practical acceptance requires a tolerance chosen for the intended application.

For two binary answers, the odds ratio (OR) is $\mathrm{OR}=p_{00}p_{11}/(p_{01}p_{10})$. OR=1 indicates independence, while values above or below 1 indicate positive or negative association under the stated coding. We use OR to assess association strength separately from marginal-share errors. Its natural logarithm places independence at zero, with larger absolute values indicating stronger association. AB and BC denote the A/B and B/C answer pairs, respectively. For the three-question cases, we report the BC odds ratio both after summing over A and within groups defined by the A answer. Separate completion has OR=1 within each A group, while mixing those groups can produce a different pooled OR.

Brier is the main comparison score. The W54/W41 matched-mean joint contrasts, OR, TV, conjunction errors, and identification bounds are post hoc diagnostics of the generated distributions. Appendix~\ref{app:selection} documents case selection and data reuse.

\subsection{Checking Compatibility Across Completion Orders}
If the conditional probabilities describe one common joint distribution, the forward and reverse completion rules must agree. Given the same A value, we compare the normalized BC tables $q_B(b|a)q_C(c|a,b)$ and $q_C(c|a)q_B(b|a,c)$. This conditional check uses model outputs alone. Population-weighted disagreement additionally uses the supplied $r(a)$, which here is an empirical human marginal. Neither check needs linked human B/C validation answers, although collecting a new order requires model calls.

Let $P$ and $Q$ be the two generated joint distributions and $D=\mathrm{TV}(P,Q)$. For any common target H, the triangle inequality gives
\begin{equation}
\max\{\mathrm{TV}(P,H),\mathrm{TV}(Q,H)\}\geq D/2.
\label{eq:orderbound}
\end{equation}
Both paths can meet a TV tolerance $\epsilon$ only if $D\leq2\epsilon$. If this condition fails, the two paths cannot both satisfy that tolerance against a common target. Linked validation records are needed to identify the more accurate path. Small disagreement can occur when both paths are accurate or when they share the same error.

\subsection{Respondent Precision and Model-Output Sensitivity}
Each request has three collected responses. We normalize vectors within the 0.01 sum tolerance and average them for the main estimates. Raw entries are reported on a 0.01 grid.

We report two kinds of uncertainty. Respondent intervals use 5000 paired resamples of survey respondents with their supplied weights, holding mean model probabilities fixed and re-estimating the reference A marginal where relevant. They describe sampling variation within a case. Output ranges recompute each result over every combination of the three collected model responses, holding people and reference shares fixed. They describe sensitivity to the collected outputs. For a triple, this gives $3^8=6561$ combinations.

Bonferroni allocation splits a family's nominal error budget across its comparisons. For a family of m contrasts, the reported percentile level is $1-0.05/m$. Appendix~\ref{app:reproduction} lists the comparison families, pointwise OR intervals, and simultaneous human-share bands. Section~\ref{sec:limitations} gives their inferential scope.

\section{Results}
Table~\ref{tab:scores} summarizes Brier gains across the evaluated cases.

\begin{table}[tbp]
\caption{Brier gains relative to the comparison-specific control. W92 uses the same mean forecast for everyone as its control. W54/W41 use population-only prediction, W29 uses prediction given recorded sex, and the triples use separate completion. Positive gains favor recorded-answer conditioning or sequential completion. The Categories column counts the scored outcomes: two answers, four answer pairs, or eight answer triples. Respondent intervals hold mean model probabilities fixed. Output ranges vary the collected model responses at fixed human records and reference shares. Gains use different controls and outcome spaces across rows.}
\label{tab:scores}
\centering\small
\begin{tabularx}{\textwidth}{P{2.5cm}r r P{3.5cm}P{3.3cm}Y}
\toprule
Comparison & Categories & Brier gain & Respondent interval & Output range & Direction across outputs \\
\midrule
W92 pairs (mean) & 2 & 0.0416 & [0.0358,0.0473] (95\%) & [0.0388,0.0433] & Positive \\
W54 pairs (mean) & 4 & 0.0153 & [0.0112,0.0193] (95\%) & [0.0123,0.0187] & Positive \\
W41 pairs (mean) & 4 & 0.0468 & [0.0356,0.0583] (95\%) & [0.0409,0.0510] & Positive \\
W29 targets (mean) & 2 & $-0.1259$ & [$-0.1772$,$-0.0737$] (95\%) & [$-0.1456$,$-0.1094$] & Negative \\
Finance1 & 8 & $-0.0013$ & [$-0.0018$,$-0.0008$] (95\%) & [$-0.0035$,0.0002] & Crosses zero \\
Finance2 & 8 & $-0.0003$ & [$-0.0026$,0.0020] (95\%) & [$-0.0027$,0.0026] & Crosses zero \\
Race & 8 & $-0.0309$ & [$-0.0389$,$-0.0232$] (97.5\%) & [$-0.0450$,$-0.0162$] & Negative \\
W29 historical context & 2 & $-0.2356$ & [$-0.3116$,$-0.1575$] (97.5\%) & [$-0.2564$,$-0.2216$] & Negative \\
\bottomrule
\end{tabularx}
\end{table}

\subsection{RQ1: Population Probabilities Can Misstate Answer Shares}
The model reproduces some population answer shares closely but substantially misstates others. Population-only predictions are uneven across the six selected requests in Table~\ref{tab:rawshares}. Retirement-account prevalence is overestimated by 13.3 pp, the public-education opinion differs by 16.0 pp, and parenting similarity differs by 23.5 pp. Medical debt and workplace similarity are closer, with errors of 2.0 and 1.9 pp. Interpreting these probabilities as population shares requires validation for the question and information condition used.

\begin{table}[tbp]
\caption{Population-only requests. Human and model columns give first-option shares (\%). Error is their absolute difference in percentage points. The table covers the six selected population-only requests.}
\label{tab:rawshares}
\centering\footnotesize
\begin{tabularx}{\linewidth}{P{0.82cm}Y rrr}
\toprule
Wave & Target & Human & Model & Error (pp) \\
\midrule
W54 & Retirement account & 56.4 & 69.7 & 13.3 \\
W54 & Medical-bill debt & 30.0 & 32.0 & 2.0 \\
W41 & Public education in 2050 & 44.7 & 28.7 & 16.0 \\
W41 & Political division in 2050 & 70.0 & 76.7 & 6.7 \\
W29 & Parenting similarity & 33.5 & 57.0 & 23.5 \\
W29 & Workplace similarity & 63.3 & 61.3 & 1.9 \\
\bottomrule
\end{tabularx}
\end{table}

\begin{figure}[tbp]
\centering\includegraphics[width=0.53\linewidth]{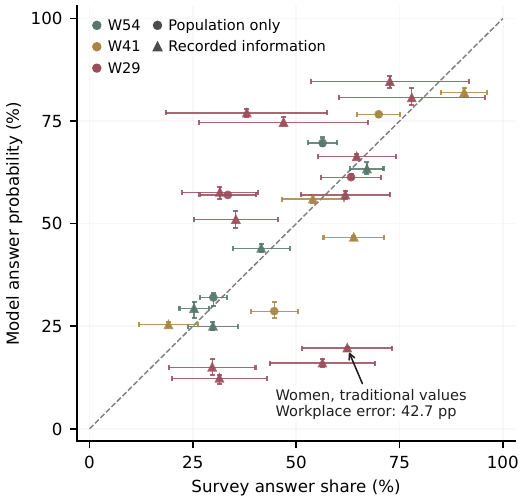}
\caption{Survey answer shares (horizontal axis) versus mean model probabilities (vertical axis) for 26 question/group/information conditions in W54, W41, and W29. Both axes refer to each question's first displayed option. Circles use population information only, while triangles add recorded respondent information. Points on the dashed diagonal match human frequencies. Horizontal bars are approximate simultaneous human-share bands under person resampling, wider than pointwise intervals. Vertical bars show the range of the three collected model outputs. The annotated women's workplace condition has the largest observed error, 42.7 pp. Both sex groups are included. Answer wording and coding appear in the item dictionary.}
\label{fig:shares}
\end{figure}

Figure~\ref{fig:shares} shows all 26 population and conditional coordinates. In the workplace item, women reporting traditional values have a human similarity share of 62.4\%, while the model gives 19.7\%, a 42.7 pp error. The large deviation is visible alongside uncertainty in human shares and the much smaller collected-output range.

\subsection{RQ2: Prediction Gains Do Not Ensure Association Fidelity}
\begin{figure}[tbp]
\centering\includegraphics[width=0.53\linewidth]{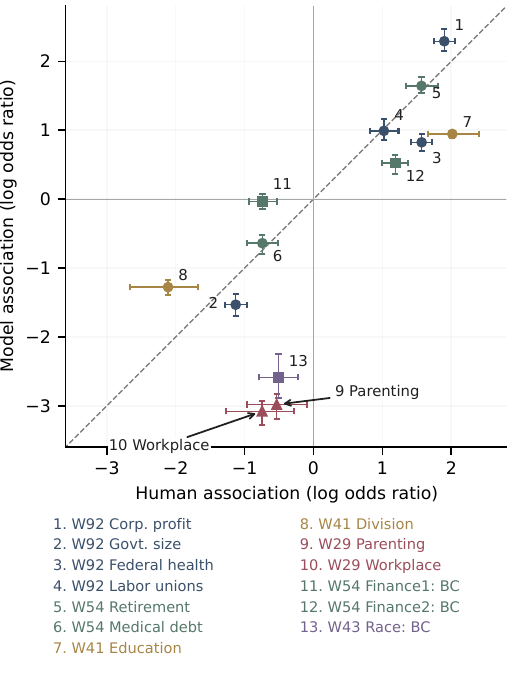}
\caption{Human association strength (horizontal axis) versus model association strength (vertical axis) for 13 answer-pair comparisons, measured by log odds ratios. Each symbol marks one comparison. The dashed diagonal indicates equal associations. Zero indicates independence. A point farther from zero vertically than horizontally has a stronger model association. Horizontal bars are pointwise nominal 95\% human respondent-bootstrap intervals. Vertical bars show the range across collected model outputs at fixed human composition/reference shares. Circles show AB pairs, triangles AB pooled over sex using empirical (X,A) composition, and squares sequential BC pooled over A. Colors identify waves. Black arrows identify 9 Parenting and 10 Workplace.}
\label{fig:or}
\end{figure}

Recorded answers can improve prediction without recovering the strength of human associations. The W92 matched-mean comparison has mean Brier gain 0.0416. The W54 and W41 anchored pair comparisons have gains 0.0153 and 0.0468. Their respondent intervals and collected-output ranges are positive for the block averages. Conditional pair completion lowers joint TV relative to population-only prediction on all four pairs under the same reference A shares. Remaining point errors are about 4--12\% (Table~\ref{tab:pairtv}). The Brier gains also remain positive against the matched-mean control, which preserves both modeled margins. Mean gains are 0.0185 for W54 and 0.0316 for W41. These comparisons support useful conditional information beyond shifts in average answer shares.

\begin{table}[tbp]
\caption{Joint TV (\%) for the four pair generators sharing an empirical A marginal. Point errors use each pair cohort.}
\label{tab:pairtv}
\centering\footnotesize
\begin{tabularx}{\linewidth}{P{0.82cm}Y rr}
\toprule
Wave & Target B & Separate & Conditional \\
\midrule
W54 & Retirement account & 13.3 & 4.7 \\
W54 & Medical-bill debt & 8.1 & 4.0 \\
W41 & Public education in 2050 & 23.4 & 12.0 \\
W41 & Political division in 2050 & 17.5 & 4.2 \\
\bottomrule
\end{tabularx}
\end{table}

Observed-answer conditioning can also produce excessive separation. In the W29 workplace item, the model predicts similarity at 16.0\% versus 80.7\% for men who say traditional values describe them well versus those who say they do not, compared with human shares of 56.3\% versus 78.0\%. Women's responses show the same pattern of excessive model separation. The appendix reports both targets for men and women.

The W29 two-target mean gain is $-0.1259$. The workplace item drives the loss, while the parenting-item gain is inconclusive. The historical-context check retains a negative gain of $-0.2356$, with a negative range across collected outputs. The prediction loss persists with an explicitly identified historical target.

Figure~\ref{fig:or} compares human and model associations for all 13 comparisons. W29 pooled human ORs of 0.59 and 0.48 become 0.051 and 0.046. W41 public-education OR falls from 7.50 to 2.58. In the W92 corporate-profit comparison, conditioning has a positive Brier gain of 0.0740 while the human OR of 6.68 becomes 9.90 in the model. Prediction gain and association fidelity can diverge within the same comparison. Across these attitude items, associations are both weakened and amplified.

Some associations are closer. Labor unions have human OR 2.78 with interval [2.27,3.43] and model OR 2.70 with collected range [2.36,3.22]. The medical-debt intervals and collected-output ranges also overlap. Simulation acceptance requires checking the relationship or event of interest alongside the prediction score.

\subsection{RQ3: Sequential Completion Does Not Ensure Joint Validity}
\begin{figure}[tbp]
\centering\includegraphics[width=0.53\linewidth]{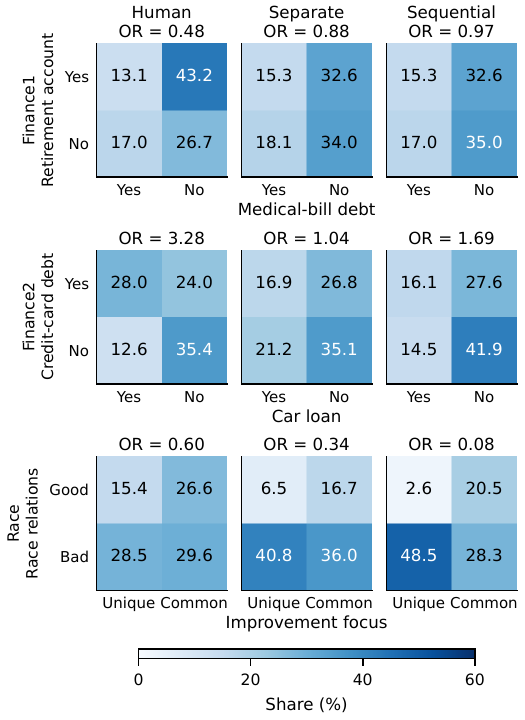}
\caption{Joint shares of answers B and C, pooled over A. Cell values are percentages. Each panel shows four combinations that total 100\% before rounding. For example, the upper-left cell of the Finance2 human panel means that 28.0\% have both credit-card debt and a car loan. Rows identify cases and columns identify human records or completion rules. Question labels apply across each row. Titles give pooled odds ratios. Both model rules use the same reference A shares. Race options are abbreviated as Unique and Common.}
\label{fig:bc}
\end{figure}

Sequential completion does not consistently improve joint validity, and changing completion order changes the generated population in both tested order comparisons. Figure~\ref{fig:bc} shows how the answer combinations differ. Both completion rules flatten Finance1's negative association. In Finance2, sequential completion strengthens an association that remains weaker than in the survey. In Race, it exaggerates the negative association.

The two financial cases do not establish a stable Brier improvement. Finance1's gain is small and changes sign across collected-output combinations. Its sequential-completion BC odds-ratio range [0.864,1.076] crosses independence, making the direction output-sensitive. Within groups defined by the A answer, human and model point estimates can also lie on opposite sides of 1. Both generated populations flatten the human association.

Finance2 illustrates why average scores and target events need separate attention. Its Brier comparison is inconclusive, and sequential completion reduces joint TV from 15.3\% to 13.5\%. Yet the share with both credit-card debt and a car loan is 16.1\%, compared with 28.0\% in the survey.

Race shows a clear loss under sequential completion. Prediction of the improvement-focus answer also worsens when the race-relations answer is observed, so the difficulty is already present before generated answers are used.

In the matched Race order comparison, reversing the completion order changes the generated population by 11.9\% of probability mass. Joint error is 30.7\% when race relations are generated first and 24.2\% when the improvement focus is generated first, compared with 23.1\% under separate completion. Both directions remain far from the survey distribution.

Finance2 also shows order dependence: its two generated populations differ by 8.8\% of probability mass. The Brier comparisons distinguish neither order from separate completion nor the two orders from one another. Order disagreement provides an internal check even when scores do not resolve a ranking. Linked validation is needed to choose the more accurate path.

Race's 11.9\% disagreement implies that at least one path has joint TV error of at least 6.0\% against any common target. Finance2's 8.8\% yields a 4.4\% bound. Their observed-output discrepancy ranges are [8.6\%,15.0\%] and [6.5\%,11.2\%]. Both cases expose disagreement across the collected combinations. Race's separate completion has 23.1\% error despite being order-invariant by construction, directly illustrating why agreement alone is insufficient.

\subsection{Sensitivity to Survey Identifiers}
Survey names and item codes may provide cues about the source of the answers. We compare named and masked requests to assess whether the findings depend on these cues. Both W54 pairs retain positive gains after survey names and item codes are masked, with positive respondent intervals and output ranges in all four comparisons. The difference between named and masked gains remains inconclusive.

Removing identifiers makes W29 workplace updating worse. The named survey produces a smaller prediction loss, and the direction of the difference persists across respondent resamples and collected outputs. Race retains negative sequential-completion gains in both identifier conditions, while the source contrast changes direction across collected outputs. Table~\ref{tab:controls} gives the paired source and order contrasts.

\begin{table}[tbp]
\caption{Five Brier control contrasts. Interval levels follow the separate families: 98.333\% for Race/W29 and 97.5\% for W54/Finance2. Collected-output ranges separately vary observed responses at fixed human data and reference shares. Positive values favor the first-named alternative.}
\label{tab:controls}
\centering\small
\begin{tabularx}{\textwidth}{P{2.35cm}Y r P{3.1cm}P{3.1cm}}
\toprule
Case family & Contrast & Estimate & Respondent interval & Output range \\
\midrule
Race/W29 & Race reverse minus forward gain & 0.0279 & [0.0206,0.0352] & [0.0204,0.0364] \\
Race/W29 & Workplace masked minus named gain & $-0.0474$ & [$-0.0581$,$-0.0359$] & [$-0.0728$,$-0.0097$] \\
Race/W29 & Race masked minus named forward gain & 0.0068 & [0.0049,0.0086] & [$-0.0091$,0.0242] \\
W54/Finance2 & W54 masked minus named mean pair gain & 0.0002 & [$-0.0023$,0.0028] & [$-0.0052$,0.0077] \\
W54/Finance2 & Finance2 reverse minus forward gain & 0.0009 & [$-0.0017$,0.0036] & [$-0.0055$,0.0069] \\
\bottomrule
\end{tabularx}
\end{table}

\FloatBarrier
\section{Discussion}
The results distinguish predictive usefulness from the validity of a generated population. We now consider what the available human information can establish and how these distinctions guide the construction and validation of synthetic respondents.

\subsection{What Human Information Can Establish}
Human margins generally bound joint frequencies without identifying a unique association. Fr\'echet bounds give the smallest and largest possible shares of a joint event consistent with the known single-question shares. They show how much remains unresolved by those shares alone. Aggregate margins, a person's known answers, and linked validation records constrain different aspects of the joint distribution. Table~\ref{tab:information} separates construction inputs from evidence about their accuracy.

\begin{table}[tbp]
\caption{Human information and the conclusions it supports. Each row describes a different information setting. Empirical shares and bounds also have sampling uncertainty.}
\label{tab:information}
\centering\small
\begin{tabularx}{\textwidth}{P{3.2cm}Y Y}
\toprule
Available information & Identified quantity or assumption & Implication in these cases \\
\midrule
No target-population human labels & Only the model's proposed distribution is observed. Frequency error cannot be measured. & Evaluation labels reveal raw errors up to 23.5 pp, which the probability vector alone does not disclose. \\
One eligible-cohort A marginal & A shares are supplied. Missing conditional frequencies remain model-dependent. & Four anchored pair generators improve but retain 4--12\% joint TV. Applicability depends on the cohort represented by the supplied A shares. \\
A person's recorded answer & Fixes this person's A answer, specifying the query $q(B|A=a)$. Estimating the corresponding human share $P(B|A=a)$ requires linked answers or additional assumptions. & W92 gains and W29 losses require separate linked-human evaluation. \\
B and C target margins & Conjunction shares lie within Fr\'echet bounds. A particular association is not identified. & Finance2 permits both-answer prevalence anywhere from 0 to 40.6\%. \\
Separate linked (A,B) and (A,C) tables & Conditional bounds are available. $B\perp C\mid A$ selects one completion through an added assumption. & Finance2's human-margin conditional-independence completion gives 21.2\%, versus the observed 28.0\%. \\
Linked validation records for A,B,C & Permit estimating joint cells, associations, event errors, and paired score differences in the evaluated cohort. & Expose Finance2's conjunction deficit and Race's distorted association. \\
\bottomrule
\end{tabularx}
\end{table}

For binary B and C, known target margins imply the classical Fr\'echet bounds
\begin{equation}
\max(0,p_B+p_C-1)\leq P(B=0,C=0)\leq\min(p_B,p_C),
\end{equation}
where $p_B=P(B=0)$ and $p_C=P(C=0)$. With compatible human (A,B) and (A,C) tables, bounds can instead be calculated within A and averaged using $P(A)$. Statistical matching's conditional independence assumption (CIA), $B\perp C\mid A$, selects $\sum_a P(a)P(B=0|a)P(C=0|a)$. This assumes no remaining B--C association within each A group and supplies a simple completion when that relationship is unobserved. The two separate tables alone cannot verify the assumption \cite{statmatchdoc}.

In Finance2, the known credit-card-debt and car-loan shares allow their overlap to range from 0 to 40.6\%. The observed overlap is 28.0\%. Conditional independence given A selects 21.2\%, illustrating how an unverified assumption can select an inaccurate point inside the feasible range. These calculations use human target margins and linked pair tables that were unavailable to the generators. Their role is to illustrate the information supplied by those inputs. Appendix~\ref{app:reproduction} gives the marginal and independence calculations.

\subsection{Checks for Constructing Synthetic Respondents}
Validation should begin with the population and output the simulation is intended to represent. Population shares test the frequency interpretation of model probabilities. Conditional shares test how a recorded answer changes the forecast, while linked answer cells test the combinations needed by the application. The cohort, period, and coding must match across these checks. Complete-case eligibility also determines which population the reference A marginal represents.

An application focused on a joint event should evaluate that event directly. Finance2 illustrates the cost of relying on an average score: sequential completion improves joint TV but underestimates the both-debt event by 11.9 pp and increases car-loan share error to 10.0 pp. Acceptance tolerances should reflect the intended use. With only marginal summaries, the bounds describe what remains unidentified and the model selects one scenario within that range. Linked records are needed to determine whether the selected relationships fit the human population.

Before linked validation answers are available, comparing completion orders can reveal internal disagreement. The order bound quantifies how far at least one path must lie from any common target, while leaving the more accurate path unidentified. Once linked records are available, event-specific errors and associations test whether either generated population fits the intended use. Reports should distinguish these empirical checks from relationships supplied by model assumptions.

\subsection{Limitations}
\label{sec:limitations}
The evidence concerns selected binary items and one model version. The cases were chosen adaptively, with two orders tested in Race and Finance2, one historical-context rewrite, and identifier controls on positive and negative cases. Estimating how often these findings occur requires a defined sample of tasks. More respondents on the same items increase within-case precision but add no new tasks.

Respondent intervals use independent-person resampling with supplied survey weights, rather than the full complex survey design. The nominal intervals omit uncertainty from adaptive selection and cannot be interpreted as between-task coverage. Output ranges summarize three collected responses per request and do not estimate future service variability. Interval/range overlap is not an equivalence test.

The target surveys are public, and the model's training exposure is unknown. Masked requests retain recognizable wording and historical dates, so they test sensitivity to explicit source identifiers. The findings establish frequency errors and incompatible generated distributions for the stated cohorts. Their mechanisms remain unobserved. Generalization to unseen surveys requires material with controlled data access.

One recorded answer per person also leaves the source of behavioral variation unidentified. A 60\% answer share could reflect stable differences between people or a 60\% response probability within each person. These cross-sectional records do not separate those explanations or the model's own uncertainty.

\section{Conclusion}
The tested probability interface shows why predictive gains, internal compatibility, and agreement with human joint answers require separate evidence. Recorded answers can help prediction while association strength remains distorted. For the tested pairs, conditional forecasts improve prediction even when both modeled margins are held fixed, with residual joint error. Sequential completion and question order also change response combinations in ways that average scores can miss. Simulation validation should target the joint events used by the application. Order checks expose internal disagreement, while human data determine whether the chosen relationships fit the population and marginal bounds show what remains unidentified.

\FloatBarrier
\section*{Acknowledgment}
C.X. was supported by Japan Society for the Promotion of Science KAKENHI grants JP23K17456, JP23K28096, JP25H01117 and JP26K03246, and Japan Science and Technology Agency CREST grant JPMJCR22M2. Z.W. was supported by Japan Science and Technology Agency BOOST grant JPMJBS2402.

\begingroup
\small
\bibliographystyle{IEEEtran}
\bibliography{references,references_web}
\endgroup

\clearpage
\appendix
\section*{Appendix}
\section{Case Selection and Data Reuse}
\label{app:selection}
Table~\ref{tab:selection} documents how the cases were selected and reused.
\begin{table}[H]
\caption{Case selection and data reuse. The cases were not sampled from a defined task population. Exploratory comparisons and controls that reuse material are identified separately.}
\label{tab:selection}
\centering\small
\begin{tabularx}{\textwidth}{P{2cm}P{5.5cm}Y}
\toprule
Case block & Selection basis & Prior exposure and interpretation \\
\midrule
W92 pairs & Four binary targets chosen by wording and availability of linked answers & Four targets evaluated under the same fixed-mean comparison. \\
W54 pairs & Savings, retirement, and medical debt chosen from related self-report items & Pair evaluation uses disjoint reference/evaluation rows. Finance1 reuses the item content. \\
W41 pairs & Related opinions in the archived 2050 scenario & Two targets evaluated as a pair block. No random domain sampling. \\
W29 predictions & Traditional values and two gender-opinion items & Recorded sex was selected after checking conditional-group eligibility and before forecasting or evaluation decoding. \\
Finance1 & Existing finance block used to test a new completion operation & Exploratory on previously analyzed item content. \\
Finance2 & Unqueried investment item and first two unqueried debt items by codebook order & Same W54 wave and split. Comparisons specified before model requests for these items. \\
Race & First three unique, complete binary racial-attitude items by public codebook wording & New wave, potentially overlapping panel people. Comparisons specified before model requests for these targets. \\
W29 historical context & Workplace item selected for its large observed prediction loss & Targeted robustness check on used people. One bundled rewrite. \\
\bottomrule
\end{tabularx}
\end{table}

The cases were selected adaptively. Finance1 is exploratory and reuses W54 pair content. The historical-context check targets W29 workplace because of its observed loss. The W29 historical-context and Race endpoints, and the two source/order contrast families, were specified before their respective model requests. Source and order controls reuse the previously evaluated people and items. Identifier-masked reverse completion is outside the evaluated design. Table~\ref{tab:alltasks} retains all 13 task comparisons.

\begin{table}[H]
\caption{All 13 reported task comparisons. Intervals are 95\% respondent-bootstrap intervals conditional on the mean probabilities. Race and the historical-context check also have the two-endpoint 97.5\% intervals. W29 rows use the original cohort description, while the historical-context check adds the 2017 field period. Some rows share questions or respondents.}
\label{tab:alltasks}
\centering\small
\begin{tabularx}{\textwidth}{Y r r r P{3.5cm}P{3.3cm}}
\toprule
Task & n & Categories & Gain & Respondent interval & Output range \\
\midrule
W92 Corporate profit & 9960 & 2 & 0.0740 & [0.0600,0.0874] & [0.0714,0.0755] \\
W92 Government size & 9958 & 2 & 0.0275 & [0.0175,0.0374] & [0.0245,0.0292] \\
W92 Federal health coverage & 10052 & 2 & 0.0402 & [0.0356,0.0451] & [0.0358,0.0437] \\
W92 Labor unions & 4891 & 2 & 0.0246 & [0.0158,0.0332] & [0.0235,0.0250] \\
W54 Retirement account & 3365 & 4 & 0.0258 & [0.0179,0.0338] & [0.0232,0.0298] \\
W54 Medical-bill debt & 3377 & 4 & 0.0048 & [0.0024,0.0072] & [0.0013,0.0076] \\
W41 Public education in 2050 & 1088 & 4 & 0.0604 & [0.0477,0.0735] & [0.0506,0.0676] \\
W41 Political division in 2050 & 1084 & 4 & 0.0332 & [0.0190,0.0476] & [0.0313,0.0344] \\
W29 Parenting similarity & 1242 & 2 & 0.0012 & [$-0.0632$,0.0649] & [$-0.0255$,0.0238] \\
W29 Workplace similarity & 1242 & 2 & $-0.2530$ & [$-0.3258$,$-0.1792$] & [$-0.2657$,$-0.2425$] \\
W54 Finance1 & 3360 & 8 & $-0.0013$ & [$-0.0018$,$-0.0008$] & [$-0.0035$,0.0002] \\
W54 Finance2 & 3360 & 8 & $-0.0003$ & [$-0.0026$,0.0020] & [$-0.0027$,0.0026] \\
W43 Race & 1612 & 8 & $-0.0309$ & [$-0.0379$,$-0.0240$] & [$-0.0450$,$-0.0162$] \\
\bottomrule
\end{tabularx}
\end{table}

\section{Association Coding and All Strata}
\label{app:association}
Tables~\ref{tab:allassoc}, \ref{tab:orstrata}, and \ref{tab:w29assoc} report overall associations, BC strata, and W29 sex-specific results, respectively. These odds ratios are post hoc descriptive diagnostics. Human quantities use original survey weights. Model quantities come from the generated joint tables. Summing over A can change an OR even when each A-stratum has conditional independence, so collapsed and conditional ratios are kept distinct. Small differences near OR=1 are not treated as established sign reversals.

\begin{table}[H]
\caption{All 13 descriptive association estimates at original mean probabilities. The first eight rows concern AB. W29 pools over sex using empirical (X,A) composition, rather than an A-only generator. The last three concern sequential BC pooled over A. Human intervals are pointwise nominal 95\% weighted person-bootstrap intervals, computed separately for each association. Model ranges vary collected outputs at fixed human composition/reference shares.}
\label{tab:allassoc}
\centering\small
\begin{tabularx}{\textwidth}{Y r P{3cm}r P{3cm}}
\toprule
Association & Human OR & Human interval & Model OR & Model output range \\
\midrule
W92 Corporate profit & 6.677 & [5.753,7.817] & 9.898 & [8.609,11.870] \\
W92 Government size & 0.324 & [0.276,0.380] & 0.216 & [0.183,0.251] \\
W92 Federal health coverage & 4.807 & [4.114,5.574] & 2.283 & [2.000,2.567] \\
W92 Labor unions & 2.780 & [2.266,3.435] & 2.698 & [2.357,3.217] \\
W54 Retirement account & 4.795 & [3.844,6.057] & 5.182 & [4.644,5.881] \\
W54 Medical-bill debt & 0.477 & [0.379,0.594] & 0.528 & [0.452,0.596] \\
W41 Public education & 7.500 & [5.258,11.088] & 2.579 & [2.425,2.660] \\
W41 Political division & 0.121 & [0.070,0.187] & 0.279 & [0.250,0.311] \\
W29 Parenting similarity & 0.587 & [0.382,0.912] & 0.051 & [0.041,0.059] \\
W29 Workplace similarity & 0.475 & [0.281,0.755] & 0.046 & [0.038,0.054] \\
Finance1 & 0.478 & [0.391,0.587] & 0.967 & [0.864,1.076] \\
Finance2 & 3.282 & [2.701,3.965] & 1.690 & [1.443,1.908] \\
Race & 0.602 & [0.452,0.797] & 0.075 & [0.056,0.106] \\
\bottomrule
\end{tabularx}
\end{table}

\begin{table}[H]
\caption{BC odds ratios and collected-output ranges. Human values include pointwise nominal 95\% person-bootstrap intervals. All-A rows collapse over A. The Separate rule has conditional OR=1 within A by construction. Output ranges are stress checks at fixed human data, not confidence intervals.}
\label{tab:orstrata}
\centering\small
\begin{tabularx}{\textwidth}{P{1.4cm}P{1.35cm}P{3.6cm}rr YY}
\toprule
Case & Stratum & Human [interval] & Separate & Sequential & Separate range & Sequential range \\
\midrule
Finance1 & All A & 0.48 [0.39,0.59] & 0.88 & 0.97 & [0.85,0.92] & [0.86,1.08] \\
Finance1 & A=0 & 0.50 [0.39,0.66] & 1.00 & 1.05 & [1.00,1.00] & [0.95,1.15] \\
Finance1 & A=1 & 0.73 [0.48,1.08] & 1.00 & 1.01 & [1.00,1.00] & [0.88,1.14] \\
Finance2 & All A & 3.28 [2.70,3.97] & 1.04 & 1.69 & [1.02,1.06] & [1.44,1.91] \\
Finance2 & A=0 & 2.74 [2.08,3.61] & 1.00 & 1.58 & [1.00,1.00] & [1.41,1.75] \\
Finance2 & A=1 & 3.70 [2.87,4.78] & 1.00 & 1.68 & [1.00,1.00] & [1.42,1.91] \\
Race & All A & 0.60 [0.45,0.80] & 0.34 & 0.08 & [0.30,0.39] & [0.06,0.11] \\
Race & A=0 & 0.93 [0.59,1.46] & 1.00 & 0.19 & [1.00,1.00] & [0.15,0.27] \\
Race & A=1 & 0.54 [0.36,0.79] & 1.00 & 0.12 & [1.00,1.00] & [0.09,0.15] \\
\bottomrule
\end{tabularx}
\end{table}

\begin{table}[H]
\caption{W29 conditioned first-option shares (\%) and descriptive ORs. Each share pair gives the values for ``describes me well and ``does not describe me well, in that order. The target first option is ``men and women are basically similar''. Both targets are shown for men and women. Human ORs include pointwise nominal 95\% person-bootstrap intervals. Model OR ranges combine the collected conditional outputs.}
\label{tab:w29assoc}
\centering\small
\begin{tabularx}{\textwidth}{P{1.9cm} P{1.1cm}P{2.4cm}P{2.4cm}P{2.7cm}r Y}
\toprule
Target & Sex & Human shares & Model shares & Human OR & Model OR & Model OR range \\
\midrule
Parenting similarity & Male & 31.4 / 46.9 & 12.3 / 74.7 & 0.52 [0.28,0.96] & 0.048 & [0.039,0.053] \\
Parenting similarity & Female & 29.7 / 38.1 & 15.0 / 77.0 & 0.69 [0.38,1.28] & 0.053 & [0.042,0.065] \\
Workplace similarity & Male & 56.3 / 78.0 & 16.0 / 80.7 & 0.36 [0.16,0.71] & 0.046 & [0.036,0.054] \\
Workplace similarity & Female & 62.4 / 72.7 & 19.7 / 84.7 & 0.62 [0.29,1.19] & 0.044 & [0.038,0.051] \\
\bottomrule
\end{tabularx}
\end{table}

\section{Item Dictionary}
\label{app:dictionary}
Tables~\ref{tab:itemdict1} and \ref{tab:itemdict2} retain all 20 question identifiers, full stems, and displayed option order. They make both the response coding and the survey cues visible. Identifiers are included for traceability rather than as extra independent observations. In the quoted retirement items, IRA means individual retirement arrangement.
\begin{table}[H]
\caption{Item wording and displayed answer coding (1/2). Identifiers are retained here for reproducibility and were present inside model-visible request content.}
\label{tab:itemdict1}
\centering\small
\begin{tabularx}{\textwidth}{P{3.5cm}Y}
\toprule
Item & Wording and options \\
\midrule
\texttt{ECONFAIR\_W92} & Please choose the statement that comes closer to your own views.\newline First: The economic system in this country unfairly favors powerful interests\newline Second: The economic system in this country is generally fair to most Americans \\
\texttt{BUSPROFIT\_W92} & Please choose the statement that comes closer to your own views.\newline First: Business corporations make too much profit\newline Second: Most corporations make a fair and reasonable amount of profit \\
\texttt{GOVSIZE1\_W92} & If you had to choose, would you rather have\newline First: A smaller government providing fewer services\newline Second: A bigger government providing more services \\
\texttt{GOVTHC\_W92} & Do you think it is the responsibility of the federal government to make sure all Americans have health care coverage?\newline First: Yes, it is\newline Second: No, it is not \\
\texttt{INSTN\_LBRUN\_W92} & Are labor unions having a positive or negative effect on the way things are going in the country these days?\newline First: Positive effect\newline Second: Negative effect \\
\texttt{FINANCEa\_W54} & Do you have a savings account?\newline First: Yes, have this\newline Second: No, do not have this \\
\texttt{FINANCEb\_W54} & Do you have an IRA, 401K or a similar kind of retirement account?\newline First: Yes, have this\newline Second: No, do not have this \\
\texttt{DEBTe\_W54} & Do you have the following type of loans or debt: Debt from medical bills\newline First: Yes, have this\newline Second: No, do not have this \\
\texttt{HAPPENc\_W41} & Still thinking ahead 30 years, which do you think is more likely to happen in the U.S.? Race relations will improve/get worse\newline First: Race relations will improve\newline Second: Race relations will get worse \\
\texttt{HAPPENf\_W41} & Still thinking ahead 30 years, which do you think is more likely to happen in the U.S.? The public education system will improve/get worse\newline First: The public education system will improve\newline Second: The public education system will get worse \\
\bottomrule
\end{tabularx}
\end{table}

\begin{table}[H]
\caption{Item wording and displayed answer coding (2/2).}
\label{tab:itemdict2}
\centering\small
\begin{tabularx}{\textwidth}{P{3.5cm}Y}
\toprule
Item & Wording and options \\
\midrule
\texttt{HAPPENj\_W41} & Still thinking ahead 30 years, which do you think is more likely to happen in the U.S.? The country will be more/less politically divided\newline First: The country will be more politically divided\newline Second: The country will be less politically divided \\
\texttt{MESUM2\_FB\_W29} & How well, if at all, do the following words or phrases describe you? Have traditional values\newline First: Describes me well\newline Second: Does not describe me well \\
\texttt{DIFF1C\_W29} & Do you think men and women are basically similar or basically different when it comes to their approach to parenting?\newline First: Men and women are basically similar\newline Second: Men and women are basically different \\
\texttt{DIFF1E\_W29} & Do you think men and women are basically similar or basically different when it comes to the things they are good at in the workplace?\newline First: Men and women are basically similar\newline Second: Men and women are basically different \\
\texttt{FINANCEc\_W54} & Do you have personal investments in stocks, bonds or mutual funds other than those held in an IRA or 401K?\newline First: Yes, have this\newline Second: No, do not have this \\
\texttt{DEBTa\_W54} & Do you have the following type of loans or debt: Credit card debt\newline First: Yes, have this\newline Second: No, do not have this \\
\texttt{DEBTb\_W54} & Do you have the following type of loans or debt: Car loan\newline First: Yes, have this\newline Second: No, do not have this \\
\texttt{RACESURV6\_W43} & When it comes to racial discrimination, which do you think is the bigger problem for the country today?\newline First: People seeing racial discrimination where it really does not exist\newline Second: People not seeing racial discrimination where it really does exist \\
\texttt{RACESURV10\_W43} & Do you think race relations in the United States are\newline First: Generally good\newline Second: Generally bad \\
\texttt{RACESURV16\_W43} & When it comes to improving race relations, do you think it is more important for people to focus on\newline First: The unique experiences of different racial and ethnic groups\newline Second: What different racial and ethnic groups have in common \\
\bottomrule
\end{tabularx}
\end{table}

\section{Reproduction and Interval Scope}
\label{app:reproduction}
The reproduction package's README gives entry commands and data-acquisition instructions. Using five externally obtained OpinionQA CSV files and cached model responses, the analyses rebuild cohort assignments, weights, scores, intervals, output ranges, associations, information bounds, and figure coordinates. The entries compare reconstructed values with the retained full-precision reference results. Figure generation uses those same numerical summaries.

\subsection{Interval Families and Collected Outputs}
Table~\ref{tab:intervalfamilies} lists the correction families. Corrections apply to the listed comparison families. Human OR intervals are pointwise nominal 95\% percentile intervals from 5000 weighted person resamples per source, sharing row draws across targets from the same source. The code uses no pseudocounts or discarded bootstrap draws. Model-output ranges hold human composition and reference shares fixed.

\begin{table}[H]
\caption{Interval families and correction scope. Brier families retain their specified contrasts. Pointwise intervals and human-share bands have separate descriptive roles. Model-output ranges summarize the collected responses at fixed human data.}
\label{tab:intervalfamilies}
\centering\small
\begin{tabularx}{\textwidth}{P{2.5cm}P{1.2cm}Y P{2.8cm}}
\toprule
Analysis family & Size & Comparisons covered & Reported level \\
\midrule
W29 historical context and Race & 2 & W29 historical-context gain and Race sequential-completion gain & 97.5\% per contrast \\
Race/W29 controls & 3 & Race reverse minus forward gain, W29 masked minus named gain, Race masked minus named forward gain & 98.333\% per contrast \\
W54/Finance2 controls & 2 & W54 masked minus named mean pair gain and Finance2 reverse minus forward gain & 97.5\% per contrast \\
Other score and OR estimates & n/a & Other original Brier comparisons, matched-mean joint diagnostics, secondary arm gains, and human odds ratios & 95\% pointwise \\
Human-share bands & 26 coordinates & Centered within-wave maxima with $0.05/3$ allocation across three waves & Approximate 95\% simultaneous bands \\
\bottomrule
\end{tabularx}
\end{table}

Across the 70 request groups used for the main comparisons and historical-context check, the median range of the first-option probability over three responses is 2 pp and the maximum is 7 pp. Finance1's joint Brier gain is $-0.0013$, with respondent interval [$-0.0018$,$-0.0008$] and output range [$-0.0035$,0.0002]. Its three complete-repeat trees have negative gains, while cross-request recombinations include a small positive gain.

The 26-coordinate human bands use a centered share bootstrap. For group indicator $g$ and first-option indicator $Y$, $\hat h_j=\sum_iw_ig_{ij}Y_{ij}/\sum_iw_ig_{ij}$. Resampled shares are centered and divided by fixed plug-in standard errors. The within-wave maximum gives a critical value at $1-0.05/3$, with Bonferroni allocation across three waves. Absolute-error bounds invert the human-share bands against fixed mean model probabilities. This uses independent-person, nondegenerate-group approximations. General process-bootstrap theory supplies background \cite{chernozhukov2016empirical}.

\subsection{Joint Prediction at Matched Margins}
The matched-mean joint comparison preserves the A and B margins of the conditional generator. W54's retirement-account and medical-debt gains are 0.0299 [0.0226,0.0373] and 0.0070 [0.0041,0.0100]. W41's public-education and political-division gains are 0.0364 [0.0285,0.0438] and 0.0268 [0.0164,0.0369]. Brackets give pointwise 95\% respondent-bootstrap intervals at fixed mean model probabilities. Wave-mean gains are 0.0185 [0.0144,0.0226] and 0.0316 [0.0242,0.0387]. Their collected-output ranges are [0.0172,0.0195] and [0.0304,0.0323]. The contrasts use all four fixed pairs, with the reference A shares and the matched mean recomputed in each respondent resample.

\subsection{Finance2 Margins and Independence Calculations}
In Finance2 the human target margins are 52.1\% credit-card debt and 40.6\% car loans. These leave the both-answer share between 0 and 40.6\%. The observed 28.0\% and model outputs 16.9\% and 16.1\% all fall inside this wide range. A point inside the bounds is feasible, not validated. The product of human margins is 21.1\%, and the human (A,B)/(A,C) CIA completion is 21.2\%. Even those stronger human inputs plus independence miss the observed conjunction. For this cell, conditioning on A does not narrow the bounds.

Those human target margins and linked tables were withheld from the original generators. These are additional-information diagnostics, not equal-budget baselines. Also, a model conjunction below the product of human margins does not alone identify association attenuation. The model's own margins differ: separate completion gives 43.7\% and 38.1\%, and sequential completion gives 43.7\% and 30.6\%. Their respective independence products are 16.6\% and 13.4\%. Association and marginal errors must be examined together.

\subsection{Secondary Log Score}
The secondary log-score gain is
\begin{equation}
G_{\log}=\sum_{a,b,c}H(a,b,c)\log\frac{q_C(c|a,b)}{q_C(c|a)},
\end{equation}
using natural logarithms and units of nats. It compares the probabilities assigned to recorded C answers with and without the recorded B answer. With shared A and B factors, it is also the joint log-score difference. Finance2's gain is 0.0132 nats, while Finance1 and Race remain negative. This secondary measure and the main Brier comparison concern different losses.

\end{document}